# Chelyabinsk meteorite explains unusual spectral properties of Baptistina Asteroid Family

Vishnu Reddy[1]
Planetary Science Institute, Tucson, AZ 85719, USA
Email: reddy@psi.edu

Juan A. Sanchez
Max Planck Institute for Solar System Research, Göttingen, Germany

William F. Bottke
Southwest Research Institute, 1050, Walnut St, Suite 300, Boulder, Co 80302, USA

Edward A. Cloutis
Department of Geography, University of Winnipeg, 515 Portage Avenue, Winnipeg, Manitoba, Canada R3B 2E9

Matthew R. M. Izawa
Department of Geography, University of Winnipeg, 515 Portage Avenue, Winnipeg, Manitoba, Canada R3B 2E9

David P. O'Brien
Planetary Science Institute, Tucson, AZ 85719, USA

Paul Mann
Department of Geography, University of Winnipeg, 515 Portage Avenue, Winnipeg, Manitoba, Canada R3B 2E9

Matthew Cuddy
Department of Geography, University of Winnipeg, 515 Portage Avenue, Winnipeg, Manitoba, Canada R3B 2E9

Lucille Le Corre
Planetary Science Institute, Tucson, AZ 85719, USA

Michael J. Gaffey
Department of Space Studies, University of North Dakota, Grand Forks, USA

Gary Fujihara
Big Kahuna Meteorites, Hilo, Hawaii, USA

Pages: 48
Figures: 11
Tables: 7

**Proposed Running Head:** Chelyabinsk, K/T Impactor, Baptistina Asteroid Family

**Editorial correspondence to:**
Vishnu Reddy
Planetary Science Institute
1700 East Fort Lowell Road, Suite 106
Tucson 85719
(808) 342-8932 (voice)
reddy@psi.edu

# Abstract

We investigated the spectral and compositional properties of Chelyabinsk meteorite to identify its possible parent body in the main asteroid belt. Our analysis shows that the meteorite contains two spectrally distinct but compositionally indistinguishable components of LL5 chondrite and shock blackened/impact melt material. Our X-ray diffraction analysis confirms that the two lithologies of the Chelyabinsk meteorite are extremely similar in modal mineralogy. The meteorite is compositionally similar to LL chondrite and its most probable parent asteroid in the main belt is a member of the Flora family. Our work confirms previous studies (e.g., Vernazza et al. 2008; de León et al., 2010; Dunn et al., 2013), linking LL chondrites to the Flora family. Intimate mixture of LL5 chondrite and shock blackened/impact melt material from Chelyabinsk provides a spectral match with (8) Flora, the largest asteroid in the Flora family. The Baptistina family and Flora family overlap each other in dynamical space. Mineralogical analysis of (298) Baptistina and 11 small family members shows that their surface compositions are similar to LL chondrites, although their absorption bands are subdued and albedos lower when compared to typical S-type asteroids. A range of intimate mixtures of LL5 chondrite and shock blackened/impact melt material from Chelyabinsk provides spectral matches for all these BAF members. We suggest that the presence of a significant shock/impact melt component in the surface regolith of BAF members could be the cause of lower albedo and subdued absorption bands. The conceptual problem with part of this scenario is that impact melts are very rare within ordinary chondrites. Of the ~42,000 ordinary chondrites, less than 0.5% (203) of them contain impact melts. A major reason that impact melts are rare in meteorites is that high impact velocities (V > 10 km/s) are needed to generate the necessary shock pressures and temperatures (e.g., Pierazzo and Melosh 1998) unless the target material is highly porous. Nearly all asteroid impacts within the main belt are at ~5 km/s (Bottke et al., 1994), which prevents them from producing much impact melt unless they are highly porous. However, shock darkening is an equally efficient process that takes place at much lower impact velocities (~2 km/s) and can cause the observed spectral effects. Spectral effects of shock darkening and impact melt are identical. The parent asteroid of BAF was either a member of the Flora family or had the same basic composition as the Floras (LL Chondrite). The shock pressures produced during the impact event generated enough impact melt or shock blackening to alter the spectral properties of BAF, but keep the BAF composition largely unchanged. Collisional mixing of shock blackened/impact melt and LL5 chondritic material could have created the Baptistina Asteroid Family with composition identical to those of the Floras, but with subdued absorption bands. Shock darkening and impact melt play an important role in altering the spectral and albedo properties of ordinary chondrites and our work confirms earlier work by Britt and Pieters (1994).

**Introduction**

*1.1    Chelyabinsk: Overview*

On February 15, 2013 a asteroid with a diameter of 17-20 meters entered the atmosphere over Chelyabinsk, Russia, and disintegrated in an airburst with an energy of ~500±100 kilotons of TNT (Brown et al., 2013). In contrast, the Tunguska event in 1908 produced an airburst with an estimated energy between 5-15 megatons (Vasilyev 1998; Boslough et al., 2008). Orbit for the meteor was calculated using video recordings of the bolide, which suggested a pre-impact orbit consistent with its origin in the inner main belt near the v6 resonance (Borovička et al., 2013). The orbit of the Chelyabinsk bolide also shows striking similarity with a Q-type (Binzel et al. 2004) near-Earth asteroid (NEA) (86039) 1999 NC43 (Borovička et al., 2013). Q-type asteroids are believed to have surface composition similar to ordinary chondrite meteorites (DeMeo et al., 2009). Other studies have suggested probable parent body for Chelyabinsk within the NEA population including 2011 EO40 (de la Fuente Marcos and de la Fuente Marcos 2013). However no direct spectral/compositional link has been made beyond the dynamical argument just like in the case of (86039) 1999 NC43. Although the airburst occurred 16 hours prior to the close flyby of a dynamically unrelated 30-meter NEA 2012 DA14, the most likely scenario is that the Chelyabinsk bolide was not connected to any known NEA. Instead, it was probably just a denizen of the overall population, which is believed to contain ~$10^7$ objects with D > 10-20 m (Rabinowitz et al., 2000; Brown et al., 2002), that happened to be in the right place at the right time to strike the Earth.

Several hundred fragments from the airburst were recovered including a 654 kg meteorite that was hoisted from the bottom of Lake Chebarkul. Laboratory analysis of recovered fragments showed that Chelyabinsk is an LL5 ordinary chondrite with olivine and low-Ca pyroxene as major mineral phases (Meteoritical Bulletin Database). The meteorite included a dark-colored fine grained shock blackened/impact melt component, which is a significant portion (~1/3) of the meteorite apart from a light-colored lithology typical of ordinary chondrites (Meteoritical Bulletin Database). LL chondrites are the least abundant of the ordinary chondrites (which include H, L and LL chondrites) comprising ~11% of observed ordinary chondrite falls (Meteoritical Bulletin Database).

Kohout et al., (2013) measured the spectra, and determined the composition, density, porosity, and magnetic susceptibility of the Chelyabinsk meteorite. They concluded that compositionally the three lithologies (shock blackened/impact melt/unaltered LL chondrite) are indistinguishable although their spectra and albedo varied. Bulk (3.32 g/cm$^3$) and grain densities (3.51 g/cm$^3$) of Chelyabinsk measured by Kohout et al., (2013) are also consistent with those of LL chondrites. The same study reported porosity values ranging from 1.5-11.4%. Unlike other LL chondrites, Chelyabinsk is reported to have more metallic iron, placing it between LL and L chondrites (Kohout et al., 2013). They also noted that the spectral effects of shock blackening and impact melt are identical.

In this work we present detailed spectral and compositional analysis of the Chelyabinsk meteorite with the goal to identify its possible parent body in the main asteroid belt. Several previous workers (e.g., Bottke et al., 2002; Vernazza et al., 2008; de León et al., 2010; Dunn et al., 2013) have suggested dynamical and compositional links between LL chondrites and the Flora family. Spectral and compositional properties of Chelyabinsk and (8) Flora, the largest member of the Flora family, were compared to identify genetic relationship between the two. Baptistina asteroid family (BAF), which is dynamically intertwined with the Flora family, has been compositionally linked to the Floras (Reddy et al., 2009, 2011). We compared spectra and composition of several BAF members with Chelyabinsk meteorite. Compositional links between Chelyabinsk and (86039) 1999 NC43 suggested by Borovička et al., (2013) is the subject of a separate but related paper.

## 2. Chelyabinsk Properties

*2.1 Laboratory Spectra: Methodology*

Spectral studies of Chelyabinsk have shown three distinct components: a) light colored lithology (LL5 chondrite), b) dark colored lithology (shock blackened LL chondrite), and c) impact melt lithology (Kohout et al., 2013). In our spectral studies we treated shock blackened and impact melt lithologies as one unit because their effect on spectra of unaltered LL chondrite material seems to be identical. Figure 1 shows the samples that were used in this study with the lighter LL5 chondrite clasts embedded in a matrix of shock blackened/impact melt. To prepare the samples for spectral analysis, the fusion crust was removed with a variable speed rotary tool and a diamond burr bit. Once free of fusion crust, the samples were crushed using alumina mortars and pestles and dry sieved to various grain sizes (<250 μm, 250-90 μm, 90-45 μm and <45 μm), whose spectra were measured. Apart from end member spectra (LL5 chondrite and impact melt), mixtures of both components were also created by crushing the larger grain size fractions to <45 μm to provide enough sample for generating mixtures.

A series of 13 mixtures involving the two end-members (Table 1) were made with the <45 μm grain size splits in succession by adding a pre-determined amount of sample to the previous mixture to minimize the amount of sample required. Samples were mixed vigorously for three minutes each prior to repacking into sample cup and collection of spectrum.

Reflectance spectra were acquired with an Analytical Spectral Devices FieldSpec Pro HR spectrometer at the HOSERLab, University of Winnipeg, over the range of 350 to 2500 nm in 1.4 nm steps, with a spectral resolution of 2 to 7 nm. The data are internally resampled by the instrument to output data at 1 nm intervals. All spectra were measured at a viewing geometry of i = 30° and e = 0° where the incident lighting was provided by an in-house 100W quartz– tungsten–halogen collimated light source. Sample spectra were measured relative to a 100% Labsphere Spectralon® standard and corrected for the minor (<2%) irregularities in its absolute reflectance and occasional detector offsets at 1000 and 1830 nm. In each case, 500 spectra of the dark current, standard, and sample were acquired and averaged, to provide sufficient signal-to-noise.

*2.2 Spectral Characteristics*

Figure 2 shows end member spectra of LL5 and shock blackened/impact melt lithologies from Chelyabinsk normalized to unity at 1.5 μm. The albedo at 0.55 μm (V band) is 9.8% for the impact melt and 27% for the LL5 chondrite material. The spectra show two absorption bands near 1 and 2 µm due to the presence of the minerals olivine and pyroxene. Mineralogical interpretation of near-IR (NIR) spectra starts with the extraction of spectral band parameters of the 1 and 2 µm pyroxene and/or olivine absorption bands (Band I and II centers, Band Area Ratio or BAR). Band centers represent the minima of continuum removed absorption bands and BAR is the ratio of the areas of Band II to Band I. Similar to albedo, the absorption bands are severely suppressed in the shock blackened/impact melt compared to unaltered lithology with band I depth of ~4% (18% unaltered) and band II depth of 2.5% (6% unaltered). Spectral slope of both lithologies is red with increasing reflectance with increasing wavelength. These spectral effects are consistent with previous studies of shock blackened ordinary chondrites (e.g., Britt and Pieters, 1994).

Continuum removed spectra of mixtures of the two lithologies are shown in Figure 3. Some mixture bins were omitted for clarity but the trends in band depth is obvious with band depth decreasing with increasing shock blackened/impact melt abundance in the mixture. Band depth drops linearly at the rate of ~1.4% for every 10% increase in the impact melt abundance. Olivine and pyroxene bands I and II centers of both LL5 chondrite and shock blackened/impact melt lithologies are indistinguishable from each other.

## 2.3 X-ray Diffraction (XRD): Methodology

Samples of the LL5 and shock blackened/impact melt lithologies were broken up and ground by hand with an alumina mortar and pestle. Large pieces of metal were removed and the powder was dry-sieved to <250 µm and <45 µm fractions. X-ray diffraction was carried out using a Bruker D8 Discover diffractometer at the HOSERLab, University of Winnipeg, using Co K$\alpha_{1,2}$ radiation ($\lambda$=1.78897 Å), accelerating voltage 40 kV and 40 mA beam current. X-ray diffraction patterns were collected for both size fractions of each lithology with step size 0.02º and counting time 10s per step covering 10º-70º 2θ. The symbol 2θ refers to the twice the Bragg angle (that is, the angle at which the incident X-ray beam is scattered as if reflected from a flat plane). The coarser <250 µm powder produced patterns qualitatively similar to that of the <45 µm fractions, but with distorted relative intensities due to large crystals. Because fine-grained material is necessary for accurate quantitative analysis using XRD and Rietveld refinement, and because the <250 µm and <45 µm fractions do not appear to be significantly different in mineralogy, the <45 µm fraction was also analyzed in a longer scan to produce Rietveld quality data, with step size 0.02º, counting time 60s per step covering 10º-70º 2θ. All subsequent discussion of XRD and Rietveld results pertains to the 60s/step 'long scan' XRD data.

### 2.3.1 Rietveld Refinement Method

Rietveld refinement uses a numerical model of a diffraction experiment, (including the incident radiation, the instrument, and the structure and composition of the scattering matter) to create a simulated diffraction pattern. The simulated diffraction pattern is then compared with the measured diffractogram for that sample, and nonlinear least-squares optimization of the model parameters (instrumental parameters, abundances and crystallographic parameters of the mineral(s) present) carried out to find the best fit of the model parameters to the diffractogram (Rietveld, 1969; Young et al., 1977; Young, 1993; Shankland, 2004). Rietveld refinement is capable of accurately quantifying the modal mineralogy of complex mixtures (e.g., Bish and Post, 1993; Gualtieri, 2000; Shankland, 2004; Wilson et al., 2006).

Rietveld refinement of X-ray diffraction patterns was performed using the BrukerAXS TOPAS version 4.2 software package. Thompson-Cox-Hastings Pseudo-Voigt functions were used to fit the diffraction peaks (Thompson et al., 1987). Because the purpose of Rietveld refinement in this study is the quantification of the mineralogy of the Chelyabinsk meteorite samples, the compositions of the minerals were not refined. Instead, the mineral models used fixed chemical compositions with average compositions reported in the literature, where available (Galimov et al., 2013; Jones et al., 2013). Starting structural models were created using published crystal structures as follows: olivine – Smyth and Hazen (1973), orthopyroxene (low-Ca pyroxene in space group Pbca) – Hugh-Jones and Angel (1994), plagioclase – Angel (1988) and Facchinelli et al. (1979), troilite – Skála et al., (2006), kamacite (Ni-bearing α-Fe) – Wilburn and Bassett (1978), and chromite – Lenaz et al., (2004). Starting structural and chemical parameters are summarized in Table 2.

*2.4 X-Ray Diffraction Results*

X-ray diffractograms for the Chelyabinsk LL5 and shock blackened/impact melt lithologies, with pattern matches from the Crystallography Online Database are shown in Figure 4. Both lithologies produce very similar diffraction patterns.

*2.4.1   Rietveld Refinement Results*

Graphical representations of the Rietveld model fit to the measured X-ray diffraction patterns for the Chelyabinsk LL5 and shock blackened/impact melt lithologies are shown in Figure 4. Refined mineral abundances and statistical goodness-of-fit parameters are reported in Table 3. Errors in the mineral abundances from the covariance matrix of the least-squares fit are less than ~0.5% (1-σ) for all refinement results reported here, though the true errors in the refined abundances may be somewhat higher. The weighted profile residual of the Rietveld refinements are all less than ~10%, and while there is no rigorous standard for the maximum $R_{wp}$ that constitutes an acceptable fit, an $R_{wp}$ value of ≤10 % is considered reasonable by many authors (Bish and Post, 1993; Young, 1993; Gualtieri, 2000; Shankland, 2004; Pecharsky and Zavalij, 2005).

*2.4.2   Mineralogy*

The mineralogy of both LL5 and shock blackened/impact melt lithologies are very similar, consisting dominantly of forsteritic olivine, orthopyroxene, with lesser plagioclase and troilite and traces of kamacite and chromite (Table 3). The shock blackened/impact melt lithology may contain slightly more troilite (6.7±0.2 vs. 4.2±0.1 wt. %) and kamacite (1.5±0.2 vs. 0.8±0.1 wt. %) and less plagioclase (8.6±0.5 vs. 11.3±0.4 wt. %). Because of the closure problem inherent in a modal mineral data set, it is more instructive to consider ratios of the measured abundances. Ratio plots of the plagioclase content to total silicate and kamacite plus troilite to total silicate (Figure 5) show a relative enrichment in troilite and metal, and a depletion in relative plagioclase content in the impact melt lithology compared to the LL5 lithology. These slight differences may reflect minor redistribution of the metal and sulfide phases and amorphization of plagioclase during shock metamorphism. Partial melting of FeS-Fe and metal-troilite assemblages have been reported for the shocked lithology of Chelyabinsk by Jones et al., (2013) and Yakovlev and Grokhovsky (2013), and partial to complete amorphization of plagioclase in Chelyabinsk impact melt was reported by Galimov et al., (2013) and in the Meteoritical Bulletin description for Chelyabinsk (Meteoritical Bulletin, no. 102, *MAPS* 48, in preparation (2014)). Other phases reported by others, e.g., taenite, ilmenite, Ca-rich clinopyroxenes, chlorapatite, merrilite, and native Cu (Galimov et al., 2013; Jones et al., 2013) were not detected and are therefore likely to be present at very low levels, less than ~0.5-1.0 wt. %. Diffractograms for both the LL5 and impact melt lithologies show similar broad features between ~12º-25º 2θ (Figure 4A, C), which may be ascribable to X-ray amorphous components, including melt glass, diaplectic glass, and nanophase materials. Because both lithologies show similar contents of X-ray amorphous materials, the presence of additional diaplectic plagioclase in the impact melt lithology may be balanced by greater concentrations of another amorphous component in the LL5 lithology. Alternatively, the sensitivity of the X-ray diffraction techniques to differences in amorphous content may not be sufficient to detect the differences between the LL5 and shock blackened/impact melt lithologies. *Despite minor differences, it is clear that the two lithologies of the Chelyabinsk meteorite are extremely similar in modal mineralogy.* Therefore, the large differences in albedo and reflectance spectra are not due to major mineralogical changes or large differences in the concentration of X-ray amorphous phases. The microstructure of the Chelyabinsk shock blackened/impact melt lithology is the most probable explanation for the differences in spectral properties, especially the presence of fine-grained metal-sulfide assemblages ('shock darkening', e.g., Rubin, 1992, Britt and Pieters 1994) in droplets, intragranular fillings, and veins (Kohout et al., 2013).

## 3.   Source Region in the Main Belt
*3.1   LL Chondrites and Flora family: Historical Review*
The Flora family has been suggested as a probable parent body of the ~6239 LL chondrite samples in terrestrial meteorite collections by several authors (e.g., Vernazza et al., 2008; de León et al., 2010; Dunn et al., 2013). Vernazza et al., (2008) noted a distinct difference in the number of NEAs with LL chondrite-like spectrum and the number of LL chondrites in the terrestrial collection. LL chondrite-like NEAs comprise about 60% of all observed NEAs whereas LL chondrite meteorites comprise only 10% of all meteorites (Vernazza et al., 2008). This discrepancy has been observed and verified by other studies (e.g., de León et al., 2010 and Dunn et al., 2013).

The reason for this difference is unknown. Both meteoroids and km-sized bodies are part of the same "collisional cascade" within the Flora family (and within the main belt in general) (see Bottke et al., 2005). Given the proximity of the Flora family to the ν6 resonance along the inner edge of the main belt region, we would expect both kinds of objects should be able to reach the ν6 resonance without major difficulty. It is possible, however, that the H and L chondrite sources are simply better at delivering meteoroids than NEAs. Much depends on the location of the H and L sources, the ages of their source bodies or families, and the flux coming from each source. If the scenario sketched here is reasonable, LLs might represent a small fraction of the total number of meteoroids, yet a larger fraction of all known NEAs.

Dunn et al., (2013) modeled the distribution of H, L and LL chondrite mineralogies of meter-sized NEAs and concluded it is comparable to NEAs in the 1-10 km range. In other words, the contribution of meteorites from meter-sized LL chondrite-like NEAs is no greater than those of their larger brethren. Using a source region model developed by Bottke et al., (2002), Dunn et al., (2013) suggested that 75% of NEAs with ordinary chondrite like composition (H, L, and LL) are preferentially derived from the ν6 resonance with a majority of them being LL chondrites (83% probability of NEAs with LL chondrites being derived from the ν6 resonance). The Flora family, which lies near the ν6 resonance has spectral properties similar to LL chondrites and is the most probable source of NEAs with this composition (Vernazza et al., 2008).

The Flora family is one the largest in the inner main belt accounting for 15-20% of all observed asteroids as of 2002 (Nesvorný et al., 2002). Their steep size-frequency distribution suggests a collisional formation (Cellino et al., 1991). The largest member of the Flora family is (8) Flora, a 140 km object that makes up nearly 80% of the total family mass. The most thoroughly studied member of the Flora family is (951) Gaspra, which was visited by the Galileo spacecraft in 1991. In an effort to identify the spectral/compositional affinity between Chelyabinsk LL chondrite and Flora family we compared their spectra and surface mineralogy.

*3.2    Spectral Match*

Spectral matching (a.k.a. curve matching) is a non-diagnostic tool that can provide general clues to the possible nature of the asteroid's surface unless there are strong diagnostic mineral absorption features (Gaffey, 2008). The spectrum of (8) Flora (Figure 6) shows strong absorption bands due to the minerals olivine and pyroxene (Gaffey, 1984). Continuum-removed spectra of laboratory mixtures of Chelyabinsk LL5 chondrite and impact melt were fitted to the telescopic spectrum to find the best match. Overlaid on the spectrum of Flora are the spectra of end members and a 50:50 mixture of shock blackened/impact melt and LL5 chondrite from Chelyabinsk that matches the observed band depth. The depth of an absorption feature is a non-diagnostic parameter that is influenced by a host of factors including: a) abundance of the absorbing species (Reddy et al., 2009); b) particle size (Reddy et al., 2012a); c) phase angle (Reddy et al., 2012a; Sanchez et al., 2012); d) abundance of opaques such as carbon and metal (Reddy et al., 2012b; Le Corre et al., 2011); impact shock and melt (Britt and Pieters, 1989,1994). To further verify this putative affinity between Flora and Chelyabinsk we conducted a mineralogical analysis of spectrum using diagnostic spectral parameters such as absorption band centers.

*3.3    Mineralogical Analysis*

Constraining the surface mineralogy of an asteroid is key to identifying its meteorite analog. Dunn et al., (2010) developed equations (Eq. 1 and 2) to derive the chemistry of olivine and pyroxene from spectra of ordinary chondrites, which have been successfully applied to derive surface composition of several asteroids (e.g., Reddy et al., 2011, 2012c; Sanchez et al., 2013, Dunn et al., 2013). Using Band I center, olivine (fayalite) and pyroxene (ferrosilite) mol. % can be estimated with uncertainties of ±1.3 and ±1.4 mol. %, respectively.

$Fa = -1284.9 * (Band\ I\ Center)^2 + 2656.5 * (Band\ I\ Center) - 1342.2$ (Eq. 1)
$Fs = -879.1 * (Band\ I\ Center)^2 + 1824.9 * (Band\ I\ Center) - 921.7$ (Eq. 2)

To verify the validity of these equations we compared the Fa and Fs values derived from the band parameters of near-Earth asteroid (25143) Itokawa, which was the target of Japanese Hayabusa sample return mission, using spectral data from Binzel et al., (2001) obtained with the NASA Infrared Telescope Facility (IRTF). We compared these spectrally derived values to those measured from samples of Itokawa returned from the Hayabusa spacecraft (Nakamura et al., 2011). Figure 7A (reproduced from Nakamura et al., 2011) shows olivine iron abundance (fayalite) on the Y-axis and pyroxene iron abundance (ferrosilite) on the X-axis from laboratory measurements of ordinary chondrites (H, L and LL), Itokawa samples returned by Hayabusa spacecraft (red filled circle), and those derived from the ground based telescopic spectrum (green filled circle). The difference between laboratory measured values of Itokawa samples and those from IRTF spectral data is less than 1 mol. % confirming Itokawa as an LL chondrite and attesting to the validity of the method we intend to apply to constrain surface mineralogy.

Similarly, we extracted the spectral band parameters for (8) Flora and used the above equations to calculate the olivine and pyroxene chemistry. Figure 7B shows a close up of the LL chondrite zone from Figure 7A along with the Fa and Fs values of (8) Flora. Flora's olivine ($Fa_{29}$) and pyroxene chemistries ($Fs_{24}$) are consistent with LL chondrites ($Fa_{27-33}Fs_{23-27}$). Laboratory measured fayalite ($Fa_{28}$) and ferrosilite values ($Fs_{23}$) values for Chelyabinsk from Meteoritical Bulletin Database (http://www.lpi.usra.edu/meteor/) are also plotted on the same figure confirming the link between the asteroid and meteorite. Chelyabinsk has relatively less iron in olivine and pyroxene than Itokawa and Flora but these differences are minor and within the uncertainties plotted. Based on the dynamical evidence from prior studies (e.g., Bottke et al., 2002; Vernazza et al., 2008; Dunn et al., 2013) and the mineralogical evidence presented here the most probable source for Chelyabinsk meteorite and other LL chondrites is the Flora family. The spectral match between Flora and 50:50 Chelyabinsk LL5 chondrite + impact melt mixture opens up the possibility of having a significant shock blackened/impact melt component in the surface regolith.

4. *Baptistina Asteroid Family: The Story so Far*
The Baptistina Asteroid Family (BAF) is located in the inner Main Belt, and was identified using the Hierarchical Clustering Method by Mothé-Diniz et al., (2005). This region is known to have other asteroid families but largely dominated by the Flora and Vesta families (Hirayama, 1918; Williams, 1992; Zappala et al., 1990, 1994, 1995; Nesvorný et al., 2005). Bottke et al. (2007) identified the Baptistina family again based on the analysis of broadband colors from the Sloan Digital Survey Moving Object Catalog (SDSS MOC) (Ivezic et al., 2001), and was later confirmed with an updated version of the SDSS MOC (Parker et al., 2008). BAF appears as a cluster in proper elements and showed a peculiar concentration of C- or X-type asteroids in a region dominated by S-types (mainly Floras). The Flora family overlaps the Baptistina family in terms of proper semimajor axis, eccentricity, inclination. In other words, it is a cluster largely within another cluster.

In 2007, Bottke et al. proposed that the catastrophic disruption of a 170 km asteroid 160+30/-20 million years ago (Myr) led to a factor of 2 increase in the terrestrial and lunar impact flux over the last ~120 My. Based on numerical modeling work, they suggested that the BAF was the most likely source of this putative increase in impactor flux. In their work, they also suggested that the K/T impactor 65 Myr was probably a former member of the BAF that had escaped into the inner solar system.

The compositional link to K/T was based on (i) the taxonomic classification using Sloan colors that suggested domination of Xc types (low albedo featureless), (ii) measurements of Cr isotopes within 65 My old layers on Earth that suggested a connection to carbonaceous chondrites, specifically a CM-type chondrite, and (iii) a 65 Myr old highly-altered fossil meteorite, presumably a fragment from the K/T impactor, that had properties similar to carbonaceous chondrite meteorites (Kyte 1998). However, spectroscopic observations of (298) Baptistina, the largest member of the family, and other family members have shown that their composition is similar to LL chondrites rather than carbonaceous chondrites (Reddy et al., 2009, 2011). This made it highly unlikely for the BAF to be the source of the K/T impactor.

Several unexplained questions remain regarding the composition of BAF that have been the source of debate in the scientific community (e.g., Masiero et al., 2011; 2013; Delbo et al., 2012). These include: a) subdued olivine and pyroxene absorption bands of BAF members compared to Floras but with surface composition similar to LL chondrites; b) X/C taxonomic classification of smaller BAF members; and c) the unusual albedo of BAF members found in the WISE dataset, with most members having albedos less than 15%. Note that Masiero et al. (2011) found a bimodal albedo distribution for the BAF in WISE, with low and high albedo components, but this probably comes from their family definition containing numerous Flora interlopers. The latest formulation in Masiero et al. (2013) appears to have corrected this problem.

Here we attempt to answer these unexplained questions using the knowledge we gained from our analysis of the Chelyabinsk LL5 chondrite and new near infrared spectral observations of BAF using the NASA IRTF.

*4.1 Observations*

Near-IR (0.7–2.5 µm) spectral observations of seven BAF members were conducted using the low-resolution SpeX instrument in prism mode (Rayner et al., 2003) on the NASA IRTF, Mauna Kea, Hawai'i between August 2011 and June 2012. Three asteroids included in this study (Baptistina, Jankovich, and 1999 VT23) were observed and analyzed as part of a previous study (Reddy et al. 2011). Apart from the asteroids, local standard stars and solar analog star observations were also performed to correct for telluric and solar continuum, respectively. Detailed description of the observing protocol is presented in Reddy (2009). Observational circumstances for SpeX data are shown in Table 4.

*4.2    Data Reduction and Analysis*

SpeX prism data were processed using the IDL-based Spextool provided by the NASA IRTF (Cushing et al., 2004). Analysis of the data to determine spectral band parameters like band centers, band depths and Band Area Ratio (BAR) was done using a Matlab code based on the protocols discussed by Cloutis et al., (1986), Gaffey et al., (2002), and Gaffey (2003, 2005). All-night average spectra were used to extract spectral band parameters. The shorter wavelength roll over for Band I pyroxene/olivine feature was set at 0.75 µm. Band I continuum was defined as a straight line between 0.75 µm and 1.5 µm. The errors of the Band I centers were estimated from ten iterations of each one of them using different order polynomial fits (typically third and fourth order), and then calculating the 1-σ (standard deviation of the mean) error from the multiple measurements of the band centers. This procedure was repeated for each all-night average spectrum and then the average 1-σ was taken as the error. We estimated 1-σ errors of 0.01 for the Band I center given the point-to-point scatter in the data.

## 5.    Chelyabinsk-Baptistina Asteroid Family Link

*5.1    Why do BAF members have subdued olivine/pyroxene bands compared to Floras?*

Reddy et al. (2009, 2011) noted that the absorption band depths of Baptistina Asteroid Family members were weaker compared to those of typical S-type asteroids. A range of factors for the cause of this difference was discussed including observational (phase angle), physical (grain size), mineralogical (mafic mineral abundance/opaques), and/or exogenic sources (carbon). While band depth is not diagnostic of mineralogy, it is an important parameter for taxonomic classification. One possible reason for the X/C taxonomic classification of BAF is due to weaker absorption bands although their mineralogy is indistinguishable from that of Floras (Reddy et al. 2011).

Similar to our spectral match of Flora in Section 3.2, we obtained spectral matches for 10 BAF members (including Baptistina) using intimate mixtures of Chelyabinsk LL5 chondrite and shock blackened/impact melt. Figures 8 shows spectra of 10 of the BAF asteroids studied overlaid with spectra of intimate mixtures that best match them. The abundance of shock blackened/impact melt material ranges from 10% (e.g., 2001 FZ63) to 100% (1998 FB147) based on the Band I depth match. Suppression of absorption bands with increasing shock blackened/impact melt abundance is due to the presence of fine-grained opaques (dominantly iron and troilite) that also lower the albedo and redden the spectral slope (Britt and Pieters 1994). Our non-diagnostic mixing experiment suggests that impact melt could suppress the absorption bands in spectra of BAF members and explain the apparent differences with those of Floras. A range of abundances of shock blackened/impact melt material on all BAF members could also potentially explain why weakly featured BAF members could easily be misclassified as X/C taxonomic types. The question then becomes whether this scenario is physically plausible (see below).

*5.2    Why do BAF members have surface mineralogy similar to Floras?*

Several previous publications (e.g., Reddy et al., 2009, 2011) have noted a similarity in surface mineralogy of BAF members and Floras. This compositional link between the two families extends to smaller members of the BAF observed as part of this study. Table 5 shows Band I centers along with olivine (fayalite) and pyroxene (ferrosilite) mol.% for all the asteroids included in this study. The ranges for LL chondrites are also listed along with those of Chelyabinsk from Kohout et al. (2013). The average fayalite ($Fa_{28}$) and forsterite ($Fs_{23}$) values for BAF members match exactly with that of Chelyabinsk ($Fa_{28}$, $Fs_{23}$) and are within the rage for LL chondrites ($Fa_{27-33}$, $Fs_{23-27}$).

We plotted the Fa and Fs values of BAF members in Figure 9 along with Flora, Baptistina, Itokawa and Chelyabinsk. The 10 BAF members we observed plot precisely on the LL chondrite region forming a continuum that spans the entire compositional range for LL chondrites. The results presented in this plot is a confirmation of our XRD work which has shown that impact melting has little or no effect on surface mineralogy. The plot also explains the apparent differences between the spectra of BAF and Flora family members but their spectrally derived compositions appear to be similar to LL chondrites (Reddy et al., 2009, 2011). Based on the similarities in surface mineralogy between the Floras and the BAF members it is plausible that the parent asteroid of the BAF was originally a member of the Floras or had surface composition similar to that of the LL chondrites.

*5.3    Why do smaller BAF members have X/C taxonomic classification?*

Several authors (e.g., Bottke et al., 2007) noted that Baptistina asteroid family is made up of distinct taxonomic types that set them apart from the background Floras. Delbo et al., (2012) noted that BAF asteroids smaller than 10 km diameter had composition consistent with X-types based on WISE albedos, Sloan colors and near-IR spectroscopic observations. X-types were originally defined under the Tholen classification system as asteroids with weakly featured/featureless spectra and no albedo information. X-types include E, M and P taxonomic types and would include a range of surface compositions from bare nickel-iron metal to primitive carbonaceous material. When albedo information is available, X-types can be differentiated as E, M and P types depending on how bright or dark they are. Masiero et al., (2011) noted that the average albedo of BAF is 0.21, which is significantly higher than most carbonaceous chondrite meteorites making them unlikely to be low albedo P taxonomic types. In Masiero et al., (2013), however, the BAF population was found to have lower albedos, with most <15%, presumably because interlopers from the Flora family had been removed. In fact, these albedos are so distinctive that one can readily separate BAF members from the background Flora population. Hence, the use of X taxonomic type for BAF would be inappropriate as we have WISE albedos for the family (Masiero et al. 2013).

We have shown that impact melt could weaken the absorption band depths and lower the overall albedo of BAF members using samples from Chelyabinsk LL5 chondrites. To further prove the limitations of applying taxonomic classification to identify surface composition of Baptistina Asteroid Family, we conducted an experiment with intimate mixtures of Chelyabinsk LL5 chondrite and shock blackened/impact melt material.

As noted earlier, the presence of different lithologies in the Chelyabinsk meteorite has a significant effect on the spectra. Figure 10 shows the NIR spectra for intimate mixtures of Chelyabinsk LL5 chondrite (LL) and shock blackened/impact melt (IM). As can be seen in this figure, increasing the fraction of the lower albedo shock blackened/impact melt component in the intimate mixture will cause a drop in reflectance throughout the entire spectrum, and a dramatic suppression of the absorption bands. From our analysis we found that the reflectance value at 0.55 μm (approximation for the geometric albedo) for the spectrum of the 100% LL chondrite sample is ~0.27, while the reflectance value at 0.55 μm for the 100% IM sample is ~0.1.

The albedo and/or intensity of the absorption bands (if present) are parameters normally used for taxonomic classification (e.g., Tholen 1984; Bus and Binzel 2002a,b; DeMeo et al. 2009). Albedo values differ among asteroids with different compositions. For example, Thomas et al., (2011), found average albedo values of ~0.29, 0.26, and 0.13 for Q-types, S-, and C-complexes, respectively among the near-Earth asteroid population. The broad range of albedos exhibited by the Chelyabinsk samples show how objects that originated in the same parent body could be classified under different taxonomic types. Thus, in order to investigate the effect of the dark impact-melt lithology on taxonomic classification, we applied the Bus-DeMeo classification (DeMeo et al., 2009) to the NIR spectra of the intimate mixtures.

The Bus-DeMeo taxonomy is based on principal component analysis and is comprised of three major complexes (S-, C-, and X-complex) and several additional classes called end members, making a total of 24 classes. It is important to point out that the Bus-DeMeo system was developed from the analysis of NIR spectra of asteroids, which typically exhibit weaker absorption bands than meteorite spectra. Therefore, caution must be taken when using this classification system with laboratory spectra of meteorite samples. In this study, however, rather than trying to find the taxonomic type of the parent body of the Chelyabinsk meteorite, we want to see whether the presence of shock darkened/impact melt material could lead to an ambiguous classification. Spectra of the intimate mixtures were classified using the online Bus-DeMeo taxonomy calculator (http://smass.mit.edu/busdemeoclass.html). With this application a spectrum is first smoothed using a cubic spline model. The smoothed spectrum is then normalized to unity at 0.55 μm, and the overall slope is removed. To this resulting spectrum the principal component analysis is finally applied. The taxonomic type assigned to each spectrum and the calculated principal components PC1' and PC2' are presented in Table 6.

Figure 11 shows a PC2' versus PC1' diagram from DeMeo et al., (2009), where the calculated PC values for the intimate mixtures are depicted as black squares. The line designated with the letter α is called "the grand divide", and it represents a natural boundary between the S-complex (and other classes like the Q-types) and the C- and X-complexes (DeMeo et al., 2009). The grand divide essentially separates objects whose spectra exhibit the 2 μm absorption band from those in which this feature is absent. Increasing PC2' values in the direction orthogonal to the line α imply that the 2 μm absorption band is becoming deeper, while the 1 μm absorption band becomes narrower (DeMeo et al., 2009). On the other hand, increasing PC1' values (parallel to the line α) implies a wider 1 μm absorption band. For the NIR spectra of the intimate mixtures we found that the sample corresponding to 100% IM is classified as either C-complex or Cb-type. For intimate mixtures with 50% < IM ≤ 95 % samples are ambiguously classified as C/Ch/Xk/Xn-type. When the intimate mixtures are composed with a fraction of IM ≤ 50% absorption bands become deeper and the spectra are classified as Q-types.

Our goal here is to demonstrate that the presence of a shock blackened/impact melt lithology in an LL chondrite matrix can lead to masking of silicate features resulting in ambiguous taxonomic classification as we suspect in the case of Baptistina Asteroid Family. While taxonomy itself remains a valuable tool for physical characterization of asteroids, it is important to account for all factors (e.g., phase angle, grain size, space weathering, impact melt, albedo) when making a asteroid-meteorite compositional link.

## 6. Implications
### 6.1 Generation of Large Scale Impact Melts

Intimate Chelyabinsk LL5 chondrite and shock blackened/impact melt provide an explanation for the observed spectral and compositional parameters of the Baptistina Asteroid Family. Our two prior works (Reddy et al., 2010 and 2011) and this work clearly demonstrate that the base composition of the BAF is similar to LL chondrites. This base composition has been modified by some mechanism(s) that manifests spectrally as weakened absorption bands and reduced absolute albedo. We have shown that shock/impact melt can cause weakening of absorption bands, lower albedo and remain compositionally indistinguishable from its parent material (LL Chondrites). The effect of shock and impact melt on spectral properties of ordinary chondrites has been explored in depth previously with similar conclusions (e.g., Britt and Pieters, 1994). They note that shock-blackened ordinary chondrites show a moderate red slope and weak absorption bands making it difficult to identify their parent bodies in the main asteroid belt. Shock blackening is largely due to the presence of finely-dispersed iron and troilite (likely formed as a result of Fe-FeS eutectic melting; the Fe-FeS eutectic is ~1261K at 1 atm) within the largely unmelted silicates (e.g., Rubin 1992; Britt and Pieters, 1994). Therefore, total impact melting is not required to induce significant changes in the optical properties of ordinary chondrites.

According to Bottke et al., (2007), Baptistina Asteroid Family was formed due to a catastrophic disruption of a 170 km asteroid. The impact velocity and the associated shock pressures for the catastrophic disruption of the parent asteroid is a key factor that affects impact melt production. Typical asteroidal impact velocities of 4-6 km/sec (Bottke et al., 1994) can generate enough shock pressure (30-35 GPa) to produce impact melts, although Hörz et al., (2005) experimentally determined that pressures >50 GPa would be needed to make the melting "optically obvious". A super-catastrophic disruption could produce shock pressures much greater than this value and experimental results suggest that a 7-8 km/sec impact could generate shock pressure excess of 80 GPa causing all solids to melt (Hörz et al. 2005). These experimental and numerical constraints must be considered in light of the well-known heterogeneity of shock response in geological materials.

Large departures from the average shock response are expected due to local variations in material properties. Impactites from terrestrial impact structures and meteorites both show evidence for extreme heterogeneities in the degree of melting and for physical mixing of materials that have experienced very different pressure-temperature histories. Regions of higher porosity, and open fissures or cracks commonly enhance melt production, as additional work is done collapsing the open space. Further, the peak shock pressures will be attenuated with distance from the point of collision. Therefore, an object formed by catastrophic disruption and later reassembly would be expected to contain a mixture of material that experienced a wide range of peak shock pressures and subsequent melting. The presence of a wide range of possible impact melt concentration in observed BAF spectra (Fig. 8 & 9) could in part reflect differences in impact melting of the protolith material, due to proximity to the point of impact and local variations in shock response, as well as collisional mixing of impact melt with chondritic material.

*6.2 Estimating the Impact Melt Abundance in Family Forming Events*
At the mean collision velocity in the asteroid belt of ~5 km/s (Bottke et al., 1994), melt production in impacts is generally found to be negligible (e.g., Keil et al., 1997). However, material porosity can significantly increase melt production in impacts (e.g., Hörz et al., 2005), and impacts above the mean velocity, while not necessarily common, can also lead to increased melt production. Davison et al. (2010) performed a suite of numerical simulations of impacts in which a spherical target impacts a body 10x its radius, for a range of impact velocities and material porosities. Vertical impacts are assumed in all cases. They then calculate the mass of material raised above the solidus temperature (given in units of impactor masses). From their results, kindly provided by the authors in tabular form, we show here several values relevant to the range of asteroidal impact velocities Table 7.

While most of the Davison et al., (2010) simulations resulted in cratering events, as opposed to catastrophic disruptions that would be more relevant to family formation, we can still use them as a basis for estimating the possible contribution of melt in the case of Baptistina family formation. If the Baptistina parent body was ~170 km diameter, a body ~40 km diameter impacting at ~5 km/s would be necessary to catastrophically disrupt the body, using best-fit strength parameters for main belt evolution from Bottke et al., (2005). That projectile has a mass of about 1% of the 150 km target. For a non-porous target, only 0.0013 impactor masses worth of melt will be produced. If the target is porous and that same impactor hit at 8-10 km/s instead of 5 km/s, giving a super-catastrophic breakup, then ~5 projectile masses worth of melt could be produced, meaning that ~5% of the material generated in the catastrophic disruption could be melt. While the question remains if 5% impact melt is enough to explain the observed spectral effect in BAF, shock blackening can be equally responsible for the observed spectral effects as there are no differences between impact melt and shock-blackening effects on spectral of Chelyabinsk (Kohout et al. 2013).

A related concern is whether or not the melt produced could be uniformly spread over the surface of most of the fragments from the impact. In an impact, melt is generated in the region a few projectile radii around the impact point, not uniformly throughout the body, so it's not clear that it could be distributed among most or all of the fragments during the post-impact phase. In addition, most of the larger family members form from the reaccumulation of smaller fragments (e.g., Michel et al., 2001), so even if the melt was evenly distributed throughout the swarm, much of it would be incorporated into the interiors of the newly-formed fragments, not just in a thin layer right on their surfaces.

*6.2 Impact Melts in Ordinary Chondrites*
While shock blackened ordinary chondrites are well documented, impact melts are relatively rare within ordinary meteorite classes. Of the 42,000 ordinary chondrites, less than 0.5% (203) of them contain impact melts. Among ordinary chondrites LL chondrites contain the least with about 21% (42) of all the OCs that have impact melts (203). In a classic review of this issue, Keil et al. (1997) discuss the numerous pitfalls in generating large impact melt volumes on asteroids. Many impact melts within ordinary chondrite meteorites are thought to be produced within impact craters on the meteorite parent body. The melt pools cool rapidly, and they tend to produce small amounts of melt that do not affect the nearby country rock. A similar problem is faced when invoking catastrophic disruption of asteroids. As Keil et al., (1997) explain, disruptive impacts do not raise the temperature of strength- or gravity-dominated asteroids by more than a few degrees. Accordingly, impact heating by single impacts is insignificant on all but the largest bodies (which can presumably retain the hottest ejecta).

A major reason impact melts are rare in meteorites is that high impact velocities (V > 10 km/s) are needed to generate the necessary shock pressures and temperatures (e.g., Pierazzo and Melosh 1998) unless the target material is highly porous. Nearly all asteroids within the main belt hit each other at ~5 km/s (Bottke et al. 1994), which prevents them from producing much impact melt. This explains why most, perhaps all asteroids families produced from high albedo parent bodies show few signs that their bodies are even marginally contaminated by impact melt. If true, where do the observed impact melts come from? Intriguingly, they may come impactors that can strike the target body from high eccentricity and inclination orbits; this places them outside the main belt region and into the planet-crossing region (Marchi et al. 2013). Such events, though, have been fairly rare over the last 3.5 Gyr.

If true, what can we make of the putative impact melt signatures on BAF members? Here is what we know:

- The family is relatively young, with an age of 90-160 My old, depending on the bulk density of the members (and how fast they drift in semimajor axis via the Yarkovksy effect) (Bottke et al., 2007; Masiero et al., 2011).
- The composition of the BAF over a range of sizes is similar to LL chondrites. But the spectra are subdued relative to typical LL chondrites or the background Flora population (Reddy et al., 2010, and 2011).
- As far as can be determined by WISE data, nearly all family members have comparable albedos (Masiero et al., 2013) and SDSS colors (Parker et al., 2008). This implies that if the dark contaminant is indeed shock blackened/impact melt material, the generation mechanism had to spread it to all family members, including those that were deep in the interior and those that were launched away at high velocities.
- As far as we know, no other family in the asteroid belt shows obvious signs for comparable pervasive shock darkening/impact melt signatures on most family members (i.e., no family identified by Masiero et al., 2013 shows moderate albedo signatures similar to that of the BAF; most are either high albedo or low albedo).
- Highly voluminous impact melt pools are fairly rare even among terrestrial and lunar craters (except in very large craters ~250 km), even though these bodies are almost always hit at V > 10 km/s.

While it is difficult to conceive a mechanism that could mix sizable fractions of impact melt in with all of the family members, shock blackening can explain some of the observed effects in BAF. The peak pressure to induce melting on decompression following an impact is well constrained (e.g., Hörz et al. 2005), however, the peak pressure required to induce shock blackening is not well know. The fact that we see shock blackened material in Chelyabinsk (shock stage S4) suggests that the peak pressures required for shock are much lower than what is needed to produce impact melt. Hence, a lot more shock-blackened material could be generated in an average impact, and the admixture of both melted and blackened material can produce the spectral changes we observe in BAF. If the parent asteroid of BAF was a member of the Floras or had the same basic composition as the Floras (LL Chondrite), and shock pressures produced during the family formation event generated enough shock blackening/impact melt material to alter the spectral properties of BAF, it could keep the BAF composition largely unchanged. Collisional mixing of shock blackened/impact melt and LL5 chondritic material could have created the Baptistina Asteroid Family with composition identical to those of the Floras, but with subdued absorption bands (Reddy et al., 2009, 2011).

6.3    *Is Chelyabinsk a sample of Baptistina Asteroid Family?*
Unlikely, but not impossible. Bischoff et al., (2013) note that the shock classification of the Chelyabinsk LL5 lithology is S4 (moderately shocked) based on the presence of crystalline plagioclase (rather than maskelynite). Estimated peak shock pressures for stage S4 chondrites are ~30-35 GPa. The shock veins and impact melt lithology likely record higher temperatures during post-shock decompression, likely due both to experiencing higher peak shock pressure and due to local heterogeneities in shock response. Hence, Chelyabinsk might be a fragment of a Flora family member, and its observed petrographic and chemical characteristics do not exclude a Baptistina Asteroid Family connection, but it is not possible to definitively link Chelyabinsk to the BAF on the basis of available data.

## 7.  Summary and Conclusions

The Chelyabinsk meteorite provided us with valuable insights into the nature of its parent asteroid in the main belt. Here we summarize the main findings of the present work:

- Chelyabinsk meteorite contains two spectrally distinct but compositionally indistinguishable components of LL5 chondritic and shock blackened/impact melt material.
- Laboratory X-ray diffraction/Rietveld refinement analysis shows that the two lithologies of the Chelyabinsk meteorite are extremely similar in modal mineralogy.
- Chelyabinsk meteorite is compositionally similar to LL chondrite and its most probable parent asteroid in the main belt is a member of the Flora family. Our work confirms previous studies (e.g., Vernazza et al. 2008; de León et al. 2010; Dunn et al. 2013) linking LL chondrites to the Flora family.
- 50:50 intimate mixture of LL5 chondrite and shock blackened/impact melt material from Chelyabinsk provides spectral match with (8) Flora, the largest asteroid in the Flora family.
- Mineralogical analysis of (298) Baptistina and 10 small Baptistina Asteroid Family members shows that their surface composition is similar to LL chondrites, although their absorption bands are subdued and albedo lower when compared to typical S-type asteroids.
- A range of intimate mixtures of LL5 chondrite and shock blackened/impact melt material from Chelyabinsk provides spectral match for all these BAF members. We suggest the presence of significant shock blackened/impact melt material in the surface regolith of BAF members could explain the lower albedo and subdued absorption bands. A problem, however, is that we so far lack a plausible physical process capable of providing shock blackened/impact melt material to all BAF members.
- Taxonomic classification of intimate mixtures of LL5 chondritic material and impact melt from Chelyabinsk using DeMeo et al. (2009) system gives ambiguous classifications ranging from C/Ch/Xk/Xn-types to Q-types depending on the abundance of shock blackened/impact melt material in the mixture.
- While taxonomy itself remains a valuable tool for physical characterization of asteroids, the presence of a shock blackened/impact melt lithology in an LL chondrite matrix can lead to ambiguous taxonomic classification as we suspect in the case of Baptistina Asteroid Family (Delbo et al. 2012).

- We hypothesize that super catastrophic disruption of an LL chondrite-like parent asteroid might have created the Baptistina Asteroid Family. Although, it is unclear if Chelyabinsk itself is a sample of the BAF.

**Acknowledgements**

This research work was supported by NASA Planetary Mission Data Analysis Program Grant NNX13AP27G, NASA NEOO Program Grant NNX12AG12G, and NASA Planetary Geology and Geophysics Grant NNX11AN84G. The authors would like to thank Nick Moskovitz and Driss Takir for their helpful reviews to improve the manuscript. VR thanks Bobby Bus for providing spectral data for (8) Flora used in this study. We thank the IRTF TAC for awarding time to this project, and to the IRTF TOs and MKSS staff for their support. EAC thanks CFI, MRIF, and CSA funding the establishment of the Planetary Spectrophotometer Facility at the University of Winnipeg, and NSERC, CSA and the University of Winnipeg for study funding.

Table 1. Weight percent abundances of the end members in the mineral mixtures that were produced from <45 µm powders of the end members.

| Sample | Shock/Impact Melt | LL Chondrite |
|---|---|---|
| CMM001 | 100% | 0% |
| CMM002 | 95% | 5% |
| CMM003 | 90% | 10% |
| CMM004 | 80% | 20% |
| CMM005 | 70% | 30% |
| CMM006 | 60% | 40% |
| CMM007 | 50% | 50% |
| CMM008 | 40% | 60% |
| CMM009 | 30% | 70% |
| CMM010 | 20% | 80% |
| CMM011 | 10% | 90% |
| CMM012 | 5% | 95% |
| CMM013 | 0% | 100% |

Table 2. Initial crystal structure and chemical models for Rietveld refinement

| Mineral | Space Group | Cell parameters (lengths in Å, angles in degrees) | Composition |
|---|---|---|---|
| Olivine | Pbnm | a=4.756 b=10.207 c=5.980 | $Mg_{1.428}Fe_{0.572}SiO_4$ ($Fa_{28.6}$) |
| Orthopyroxene | Pbnm | a=18.23 b=8.819 c=5.1802 | $Ca_{0.016}Mg_{0.749}Fe_{0.235}SiO_3$ ($Fs_{23.5}Wo_{1.6}$) |
| Plagioclase | P$\bar{1}$ | a=8.183 b=12.883 c=14.186 α=93.38 β=115.87 γ=90.82 | $Ca_{0.085}Na_{0.845}K_{0.070}Al_{1.085}Si_{2.915}O_8$ ($An_{8.5}Ab_{84.5}Or_7$) |
| Troilite | P$\bar{6}$2c | a=5.9650 c=11.7570 | FeS |
| Kamacite | Im$\bar{3}$m | a=2.866 | $Fe_{0.93}Ni_{0.05}Co_{0.02}$ |
| Chromite | Fd$\bar{3}$m | a=8.33273 | $Fe_{0.90}Mg_{0.10}Cr_{1.90}Al_{0.10}O_4$ |

Table 3: Rietveld refinement results

| Mineral | LL5 | Shock/Impact melt |
|---|---|---|
| Olivine | 56.2 ± 0.5 | 57.3 ± 0.6 |
| Orthopyroxene | 26.9 ± 0.4 | 25.3 ± 0.6 |
| Plagioclase | 11.3 ± 0.4 | 8.6 ± 0.5 |
| Troilite | 4.2 ± 0.1 | 6.7 ± 0.2 |
| Kamacite | 0.8 ± 0.1 | 1.5 ± 0.2 |
| Chromite | 0.6 ± 0.1 | 0.7 ± 0.1 |
| $R_{wp}$ (%) | 9.18 | 9.48 |
| Plag/Total Silicate | 0.12 ± 0.01 | 0.09 ± 0.02 |
| (Kam+Tro)/Total Silicate | 0.05 ± 0.01 | 0.09 ± 0.01 |

Concentrations in wt.%. Uncertainties (1-σ) are derived from the covariance matrix of the Rietveld refinement fit, and may underestimate the true uncertainties. Uncertainties (1-σ) in abundance ratios are calculated assuming zero covariance.

Table 4. Observational circumstances for Baptistina Family Asteroids included as part of this study.

| Asteroid | Observation Date | V. Magnitude | Phase Angle |
|---|---|---|---|
| (298) Baptistina | March 21, 2008 | 13.8 | 14.1 |
| (6589) Jankovich | October 20, 2009 | 16.3 | 6.2 |
| (21910) 1999 VT23 | September 14, 2009 | 15.2 | 26.8 |
| (25985) 2001 FZ63 | June 24, 2012 | 17.8 | 15.9 |
| (30635) 1997 JZ12 | August 26, 2011 | 17.7 | 11.2 |
| (88276) 2001 MN6 | August 27, 2011 | 17.9 | 12.6 |
| (88699) 2001 RW142 | August 27, 2011 | 17.7 | 5.1 |
| (124522) 2001 RN77 | August 28, 2011 | 17.5 | 1.8 |
| (137222) 1999 RB1 | June 25, 2012 | 17.9 | 14.7 |
| (113032) 2002 RH48 | June 25, 2012 | 17.8 | 10.3 |

Table 5. Physical, spectral and mineralogical parameters for asteroids and meteorites as part of this study. The olivine and pyroxene mol. % were calculated using Eq. 1 and 2; the absolute magnitude (H) is from the Minor Planet Center; and the diameter for Baptistina Asteroid Family members was calculated using equation from Fowler and Chillemi (1992) based on average family albedo of 0.21 from Masiero et al. (2011). Itokawa diameter is the long axis as measured by the Hayabusa mission and Flora diameter is measured from adaptive optics. Chelyabinsk fayalite and ferrosilite values are from Meteoritical Bulletin Database.

| Asteroid | Band I Centers | Olivine (Fayalite) | Pyroxene (Ferrosillite) | *H* Magnitude | Diameter |
|---|---|---|---|---|---|
| | (µm) | (mol. %) | (mol. %) | | km |
| (8) Flora | 1.00 | 29 | 24 | 6.49 | 125.0 |
| (298) Baptistina | 1.00 | 29 | 24 | 11.32 | 18.68 |
| (6589) Jankovich | 0.98 | 27 | 22 | 14.6 | 4.125 |
| (21910) 1999 VT23 | 0.97 | 26 | 21 | 13.6 | 6.538 |
| (25143) Itokawa | 0.99 | 28 | 23 | 19.2 | 0.535 |
| (25985) 2001 FZ63 | 0.96 | 24 | 20 | 15.3 | 2.988 |
| (30635) 1997 JZ12 | 1.01 | 30 | 25 | 15.8 | 2.373 |
| (88276) 2001 MN6 | 0.97 | 26 | 21 | 15.8 | 2.373 |
| (88699) 2001 RW142 | 1.00 | 29 | 24 | 15.8 | 2.373 |
| (124522) 2001 RN77 | 0.98 | 27 | 22 | 16.0 | 2.165 |
| (113032) 2002 RH48 | 0.98 | 27 | 22 | 15.9 | 2.267 |
| (137222) 1999 RB1 | 1.03 | 31 | 25 | 15.9 | 2.267 |
| | | | | | |
| Chelyabinsk | | 28 | 23 | | |
| LL Chondrites | | 27-33 | 23-27 | | |
| Average BAF | | 28 | 23 | | |

Table 6. Intimate mixtures: unaltered LL chondrite (LL) + shock/impact-melt (IM). The columns in this Table are: sample, taxonomic classification (Bus-DeMeo), and principal components PC1' and PC2'.

| Sample | Taxonomy | PC1' | PC2' |
|---|---|---|---|
| 100% LL | Q-type | −0.6999 | 0.2477 |
| 100% IM | C-complex, Cb-type | −0.5738 | 0.0255 |
| 95% IM 5% LL | C/Ch/Xk/Xn-type | −0.5637 | 0.0208 |
| 90% IM 10% LL | C/Ch/Xk/Xn-type | −0.5774 | 0.0416 |
| 80% IM 20% LL | C/Ch/Xk/Xn-type | −0.6146 | 0.0622 |
| 70% IM 30% LL | C/Ch/Xk/Xn-type | −0.6251 | 0.0859 |
| 60% IM 40% LL | C/Ch/Xk/Xn-type | −0.6342 | 0.1022 |
| 50% IM 50% LL | Q-type | −0.6517 | 0.1403 |
| 40% IM 60% LL | Q-type | −0.6432 | 0.1442 |
| 30% IM 70% LL | Q-type | −0.6744 | 0.1807 |
| 20% IM 80% LL | Q-type | −0.6674 | 0.1891 |
| 10% IM 90% LL | Q-type | −0.6739 | 0.2209 |

Table 7. Results of a suite of numerical simulations of impacts performed by Davison et al. (2010) using a spherical target impacts a body 10x its radius, for a range of impact velocities and material porosities. Vertical impacts are assumed in all cases. The mass of material raised above the solidus temperature (given in units of impactor masses) is shown for a range of porosities and impact velocities.

| No Porosity | | 20% Porosity | | 50% Porosity | |
|---|---|---|---|---|---|
| Velocity (km/s) | Mass of Material Above Solidus Temperature (units of impactor mass) | Velocity (km/s) | Mass of Material Above Solidus Temperature (units of impactor mass) | Velocity (km/s) | Mass of Material Above Solidus Temperature (units of impactor mass) |
| 5 | 0.0013 | 5 | 0.15 | 5 | 1.7 |
| 8 | 0.12 | 8 | 2.8 | 8 | 4.9 |
| 10 | 0.76 | 10 | 6.7 | 10 | 7.2 |

**Figure Captions**

Figure 1. Sample of Chelyabinsk LL5 chondrite that was used in this study with the lighter LL5 chondrite clasts embedded in a matrix of shock blackened/impact melt material.

Figure 2. End member spectra of LL5 chondrite and shock blackened/impact melt lithologies from Chelyabinsk normalized to unity at 1.5 µm. The spectra show two absorption bands and 1 and 2 µm due to the presence of the minerals olivine and pyroxene.

Figure 3. Continuum removed spectra of mixtures of LL5 chondrite and shock blackened/impact melt lithologies. Some mixture bins were omitted for clarity but the trends in band depth is obvious with band depth decreasing with increasing shock/impact melt abundance in the mixture.

Figure 4. X-ray diffractograms for the Chelyabinsk LL5 and shock blackened/impact melt lithologies, with pattern matches for olivine, orthopyroxene, plagioclase, troilite, kamacite and chromite from the Crystallography Open Database (Grazulis et al., 2009). Diffractograms for both LL5 and shock blackened/impact melt lithologies are very similar. A) Pattern matches for the Chelyabinsk LL5 lithology. B) Rietveld refinement results for the Chelyabinsk LL5 lithology, data in blue, Rietveld model in red, residual in grey, small markers below denote the angular positions of diffraction peaks. C) Pattern matches for the Chelyabinsk shock blackened/impact melt lithology. D) Rietveld refinement results for the Chelyabinsk shock blackened/impact melt lithology, data in blue, Rietveld model in red, residual in grey, small markers below denote the angular positions of diffraction peaks.

Figure 5. Ratio plots of the plagioclase content to total silicate and kamacite plus troilite to total silicate indicating a relative enrichment in troilite and metal, and a depletion in relative plagioclase content in the shock blackened/impact melt lithology compared to the LL5 lithology.

Figure 6. IRTF spectrum of (8) Flora showing strong absorption bands due to the minerals olivine and pyroxene (DeMeo et al. 2009). Continuum removed spectra of laboratory mixtures of Chelyabinsk LL5 chondrite and impact melt were fitted to the telescopic spectrum to find the best match. Overlaid on the spectrum of Flora are the spectra of end members and a 50:50 mixture of shock blackened/impact melt and LL5 chondrite material from Chelyabinsk that matches the observed band depth.

Figure 7. (A) Plot showing olivine iron content (mol.% fayalite) on the Y-axis and pyroxene iron content (mol.% ferrosilite) on the X-axis from laboratory measurements of ordinary chondrites (H, L and LL), Itokawa samples returned by Hayabusa spacecraft (red filled circle), and those derived from the ground based telescopic spectrum (blue filled square). The difference between laboratory

measured values of Itokawa samples and those from IRTF spectral data is less than 1 mol. % confirming Itokawa as an LL chondrite and attesting to the validity of the method we intend to apply to constrain surface mineralogy. Figure 7B shows a close up of the LL chondrite zone from Figure 7A along with the Fa and Fs values of (8) Flora. Flora's olivine (Fa29) and pyroxene chemistries (Fs24) are consistent with LL chondrites (Fa27-33Fs23-27). Laboratory measured fayalite (Fa28) and ferrosilite values (Fs23) values for Chelyabinsk from Meteoritical Bulletin Database are also plotted on the same figure confirming the link between the asteroid and meteorite. The data for ordinary chondrites is from Nakamura et al. 2011.

Figure 8. Plot showing spectral match for 10 BAF members (30635) 1997 JZ12, 2001 FZ63, 1999 RB1, 2001 MN6, 2001 RN77, 2001 RW142, (6589) Jankovich, 1999 VT23, (298 Baptistina), and 2002 RH48; and intimate mixtures of Chelyabinsk shock blackened/impact melt and LL5 chondrite material. Spectra of the BAF asteroids overlaid with spectra of intimate mixtures that best match them. In this Figure, the abundance of impact melt ranges from 10% (e.g., 2001 FZ63) to 70% (e.g., 6589 Jankovich) based on the Band I depth match. Suppression of absorption bands with increasing shock/impact melt abundance is due to the presence of fine-grained opaques that also lower the albedo and redden the spectral slope.

Figure 9. Plot showing fayalite and ferrosillite mol. % of BAF members along with Flora, Baptistina, Itokawa and Chelyabinsk. Flora and Baptistina have the same spectrally-derived fayalite and ferrosillite mol. % and hence Flora is masked behind the Baptistina data point. The 10 BAF members we observed plot precisely on the LL chondrite region forming a continuum that spans the entire compositional range for LL chondrites. The results presented in this plot is a confirmation of our XRD work which shows that impact melt has the same composition as its parent material.

Figure 10. Near-IR spectra for intimate mixtures of Chelyabinsk LL5 chondrite (LL) and shock/impact melt (IM). Increasing the fraction of the lower albedo shock darkened/impact melt material in the intimate mixture will cause a drop in reflectance throughout the entire spectrum, and a dramatic suppression of the absorption bands. From our analysis we found that the reflectance value at 0.55 μm for the spectrum of the 100% LL sample is ~ 0.27, while the reflectance value at 0.55 μm for the 100% IM sample is ~ 0.1.

Figure 11. Plot showing the principle components (PC2' versus PC1' diagram) from DeMeo et al. (2009), where the calculated PC values for the intimate mixtures are depicted as black squares. The line designated with the letter α is called "the grand divide", and it represents a natural boundary between the S-complex (and other classes like the Q-types) and the C- and X-complexes (DeMeo et al., 2009). The grand divide essentially separates objects whose spectra exhibit the 2 μm absorption band from those in which this feature is absent. Increasing PC2' values in the direction orthogonal to line α implies that the 2 μm absorption band is becoming deeper, while the 1 μm absorption band becomes narrower (DeMeo et al., 2009).

Figure 1.

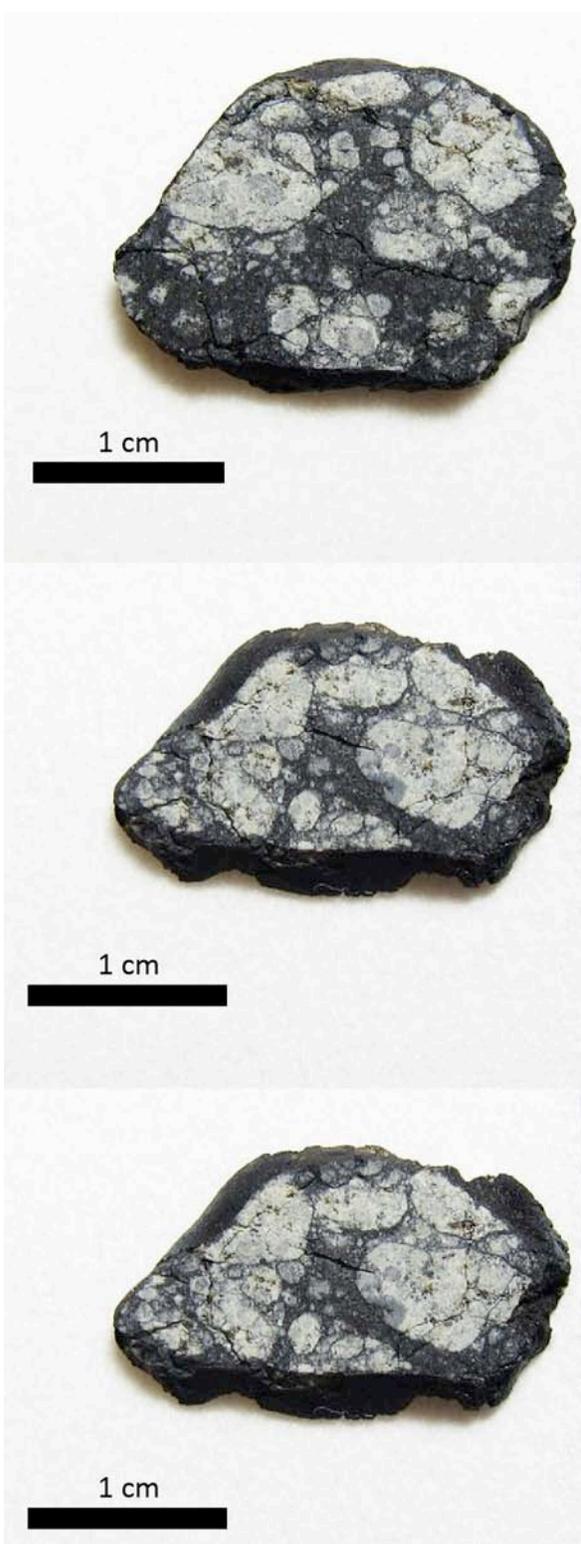

Figure 2.

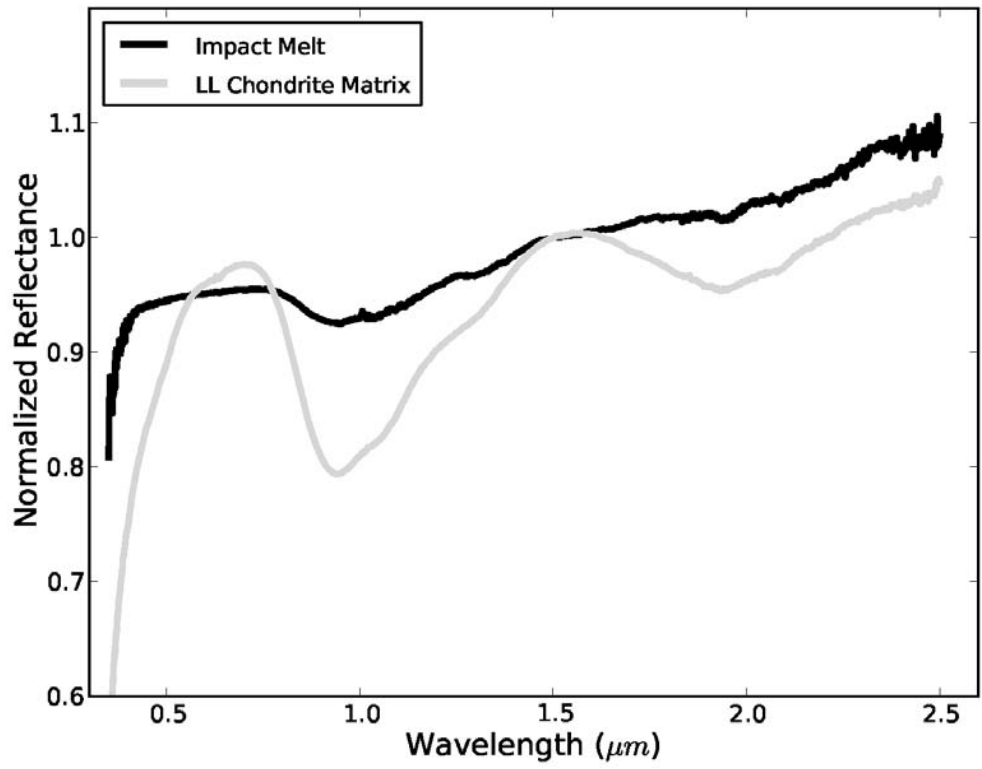

Figure 3.

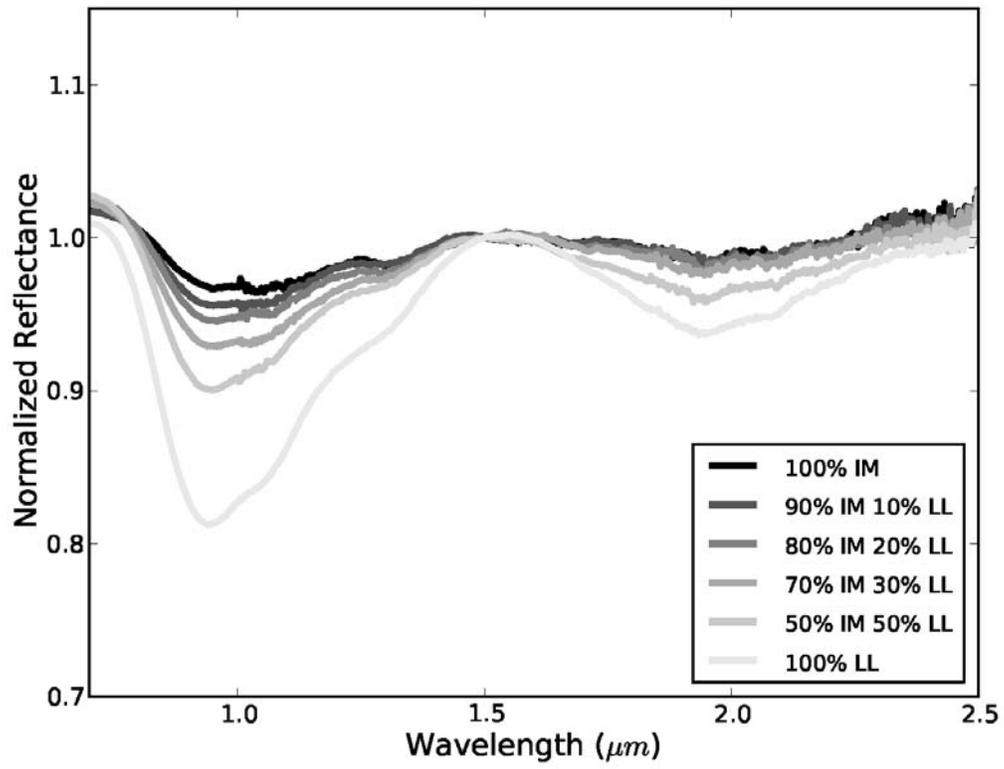

Figure 4.

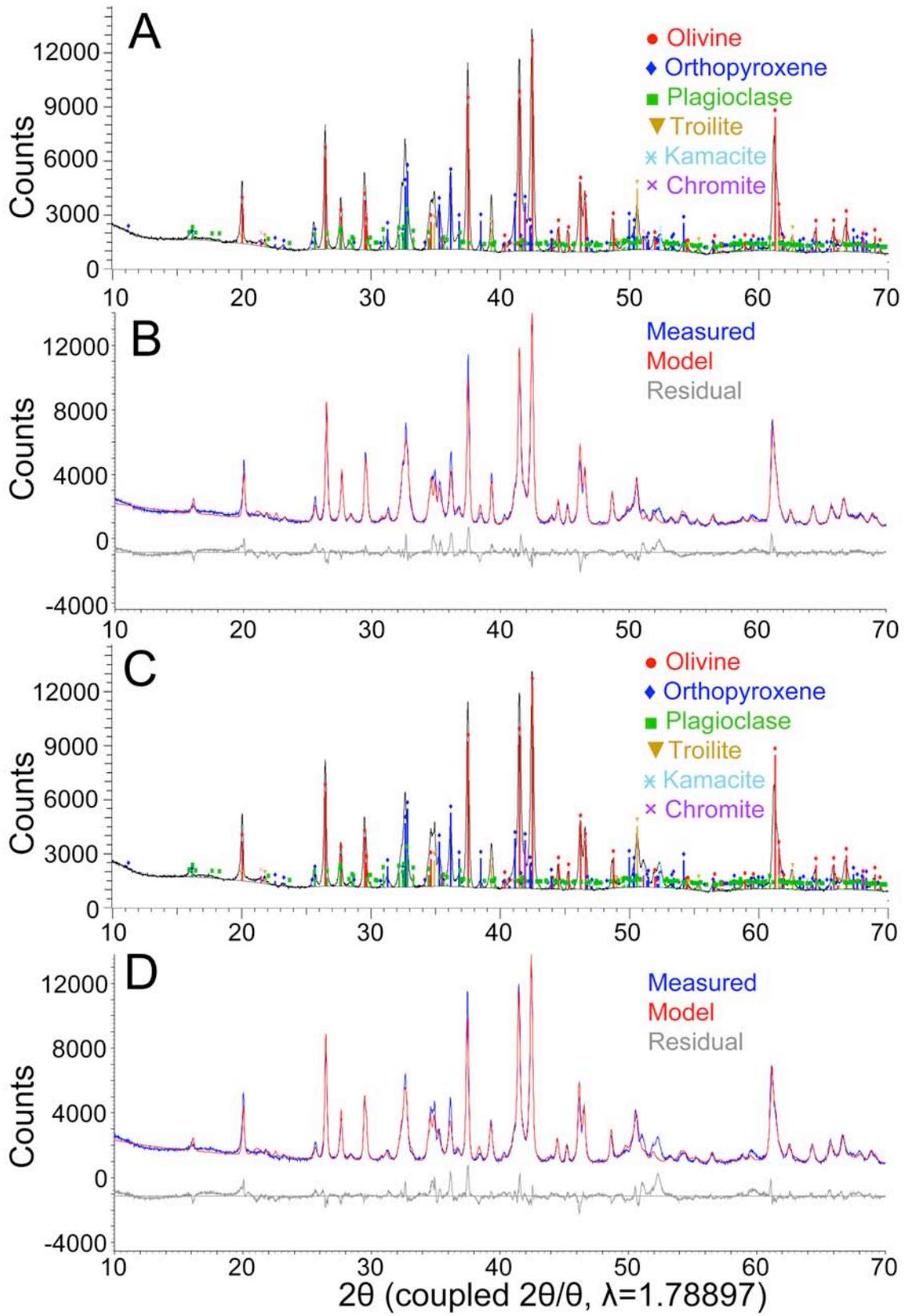

Figure 5.

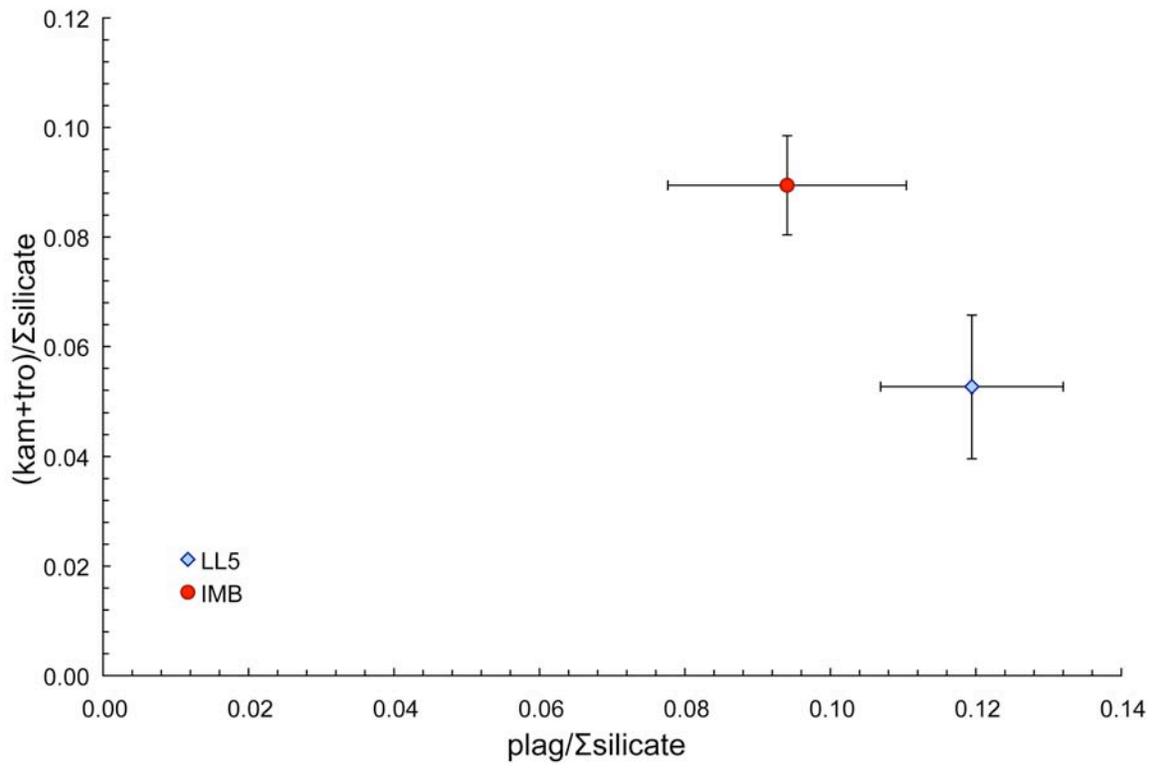

Figure 6.

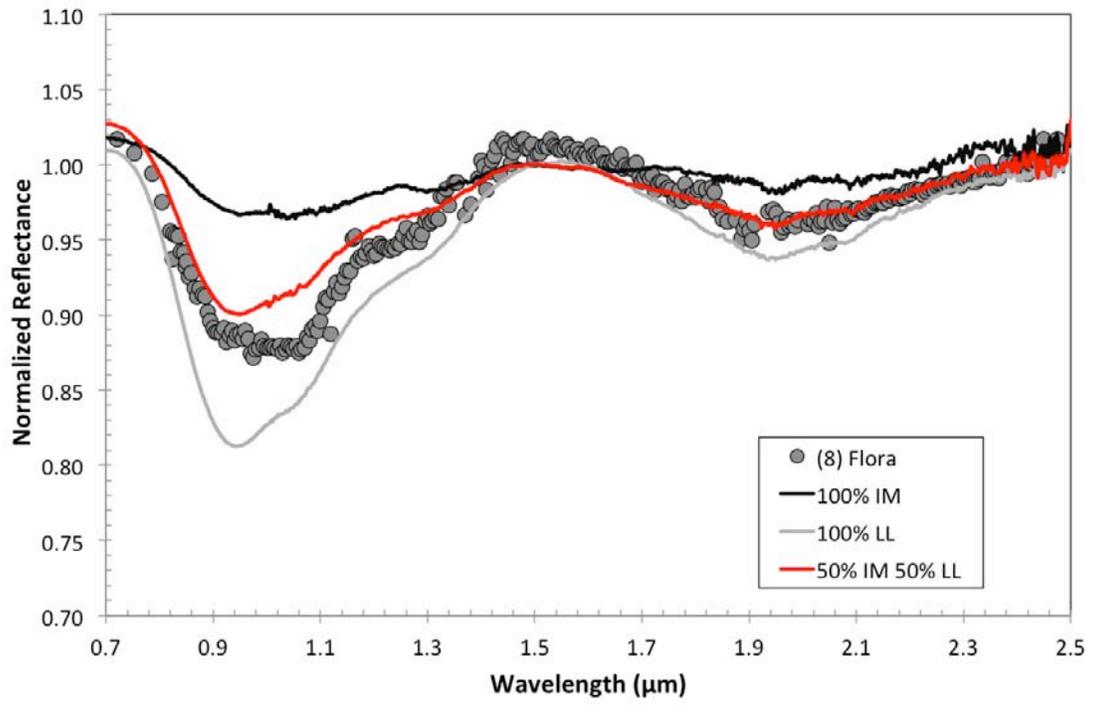

Figure 7.

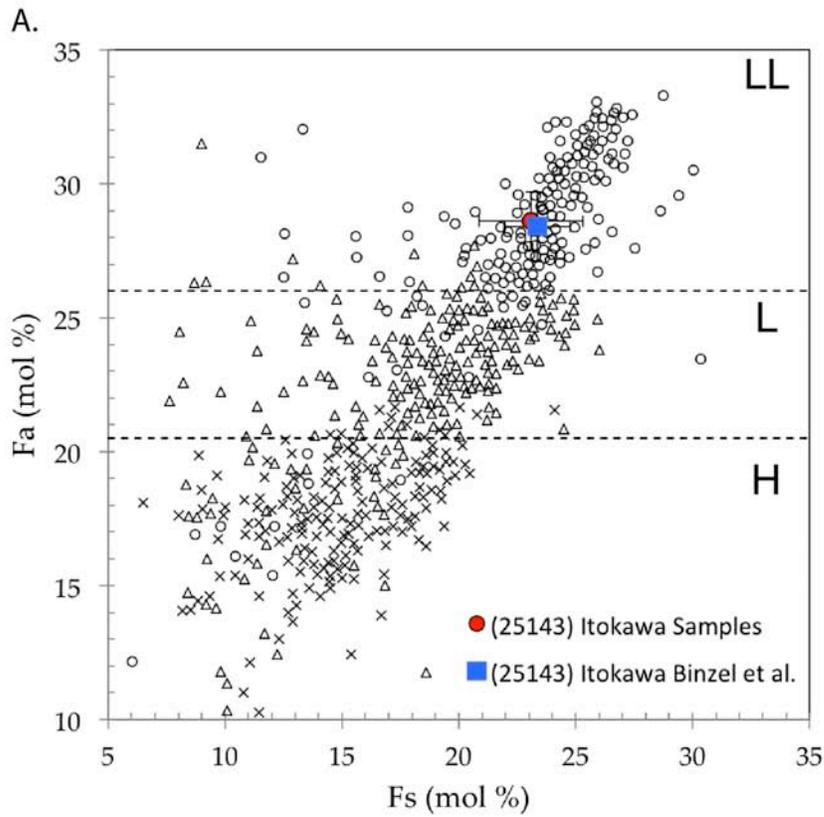

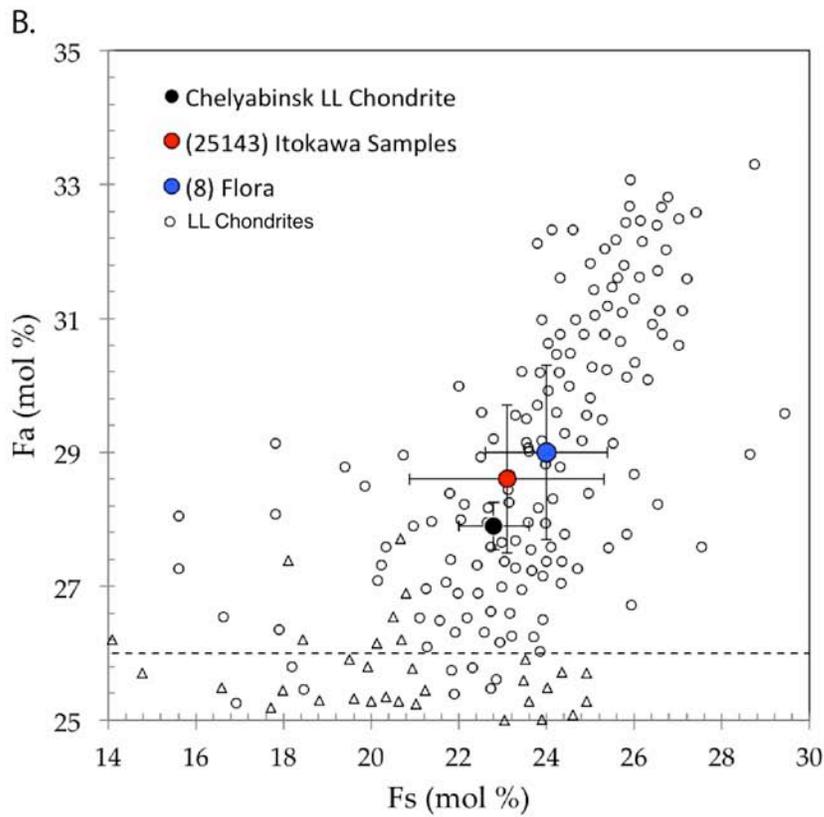

Figure 8.

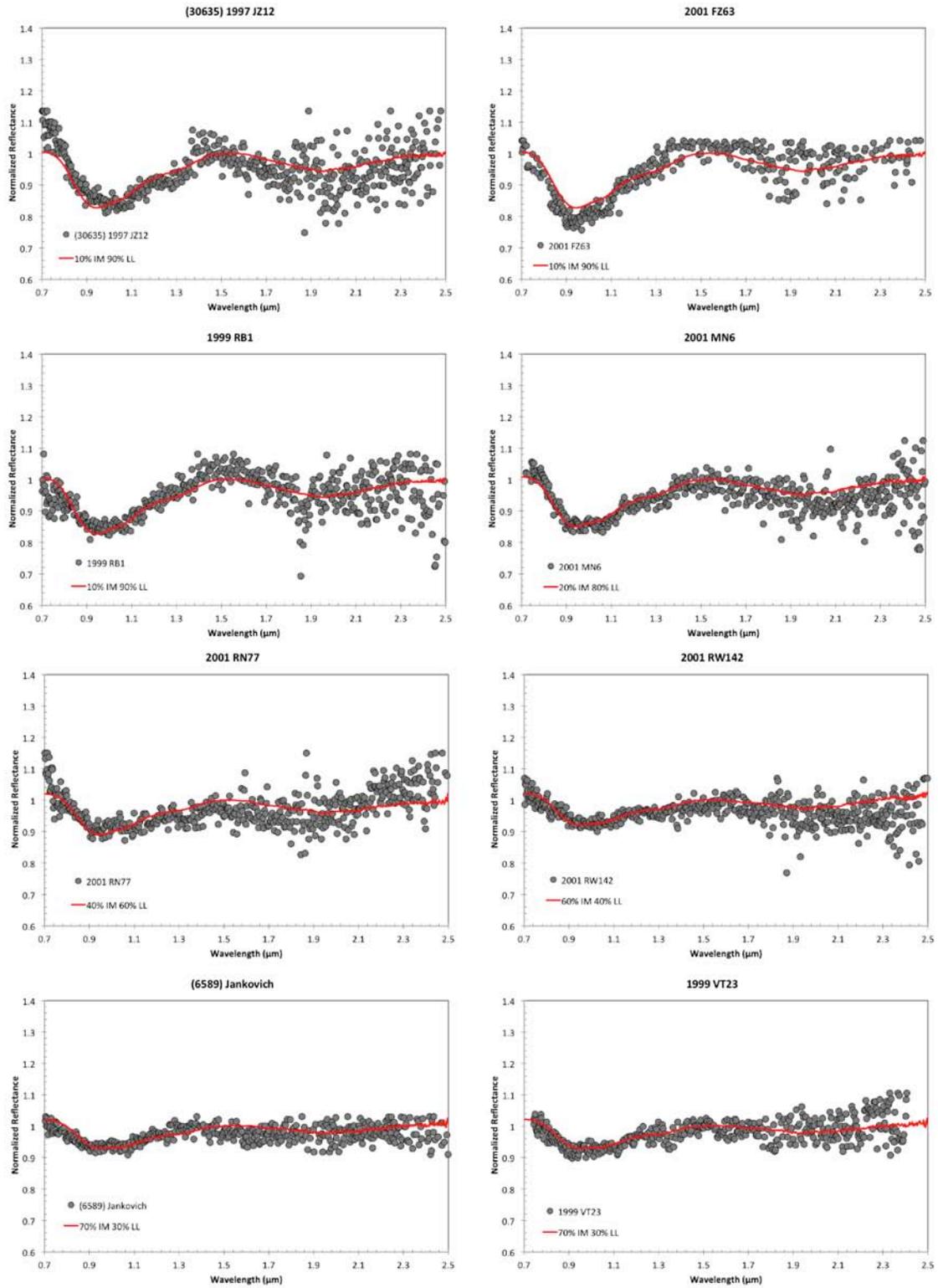

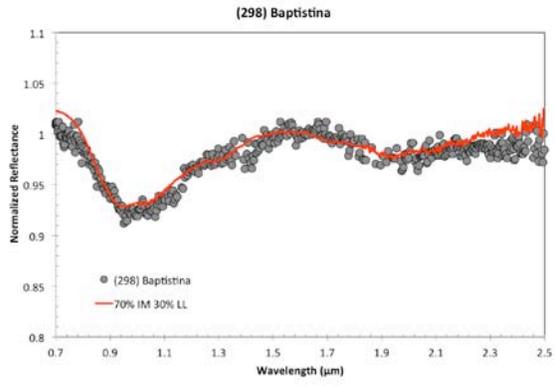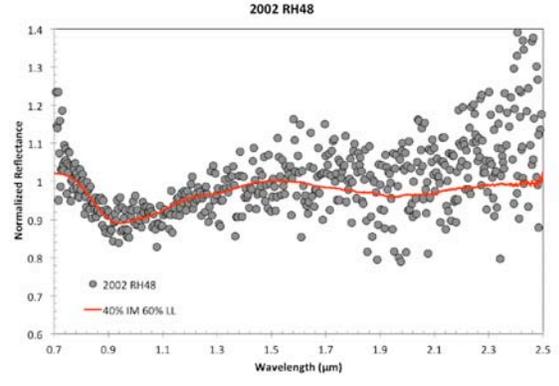

Figure 9

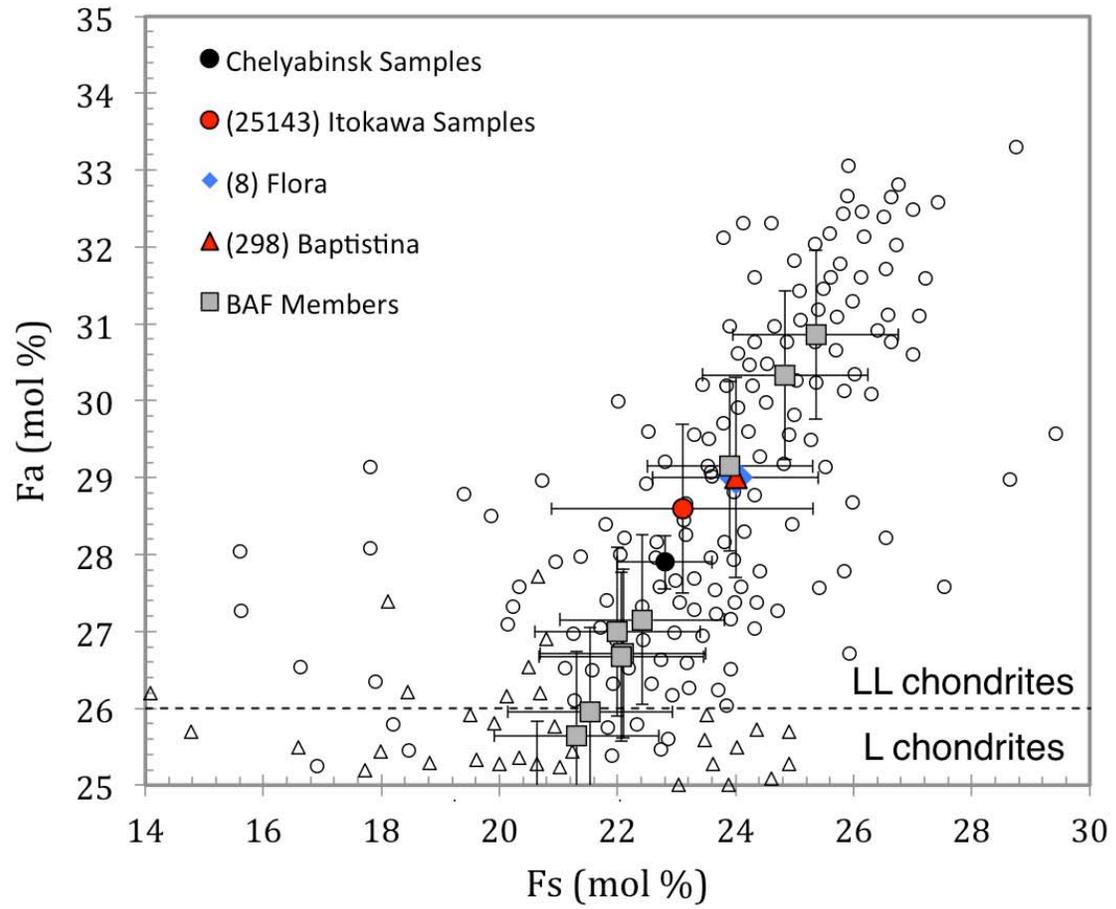

Figure 10.

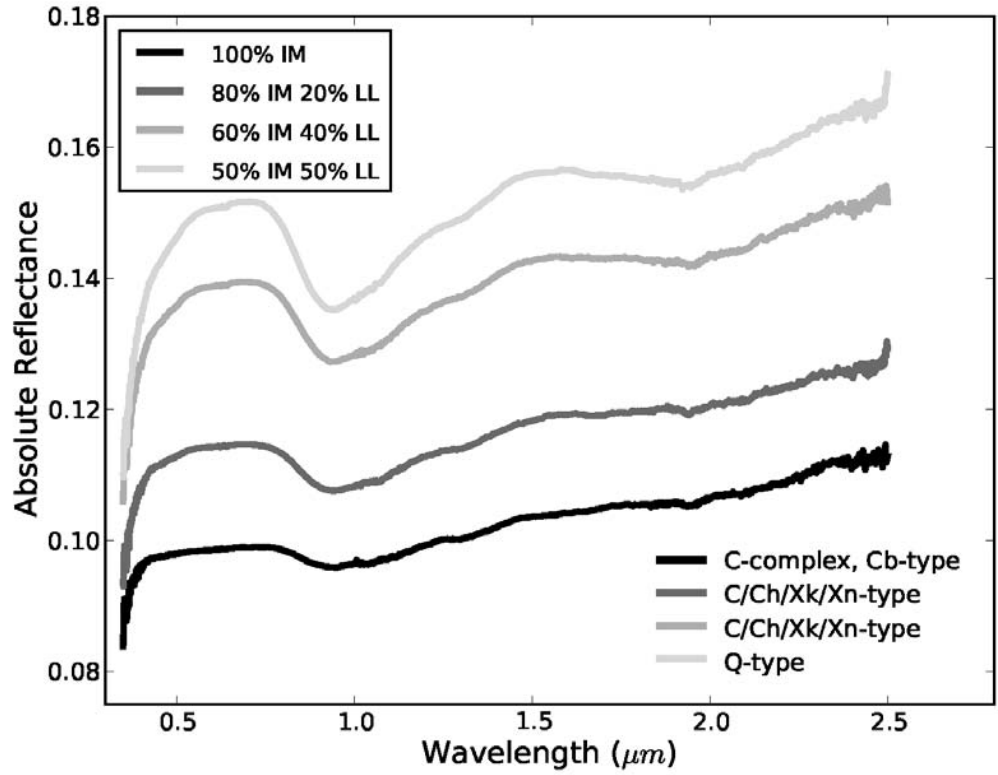

Figure 11.

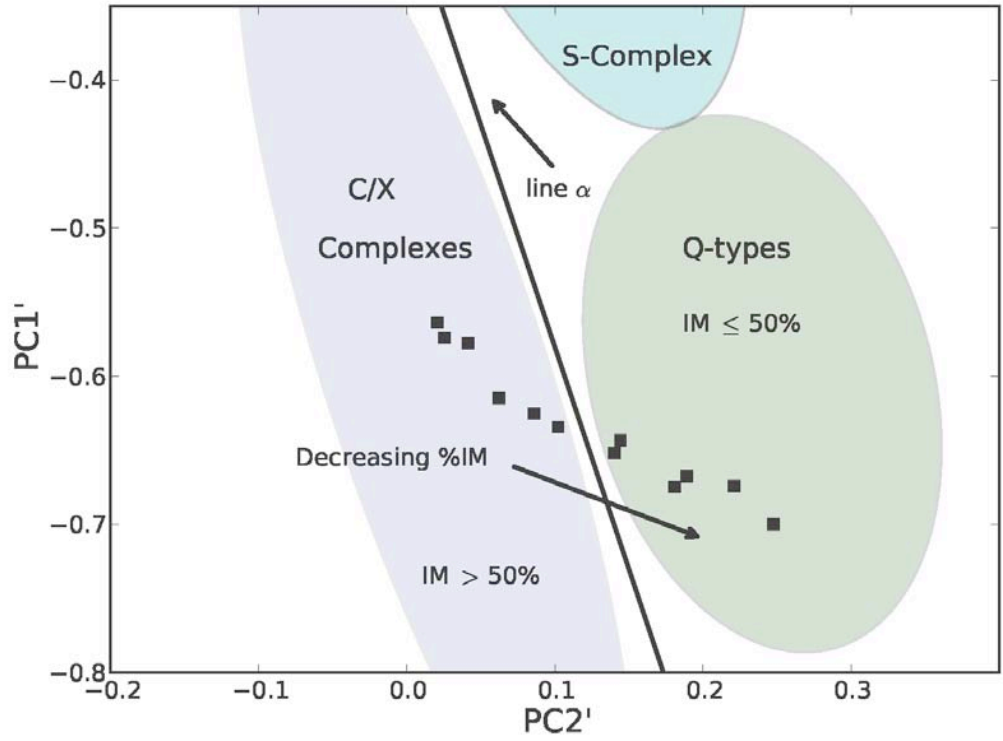